\documentclass[preprint,amsmath,amssymb,aps,prstab]{revtex4-1}%

\usepackage{dcolumn}
\usepackage{bm}
\usepackage{siunitx}
\usepackage{pifont}
\newcommand{\cmark}{\ding{51}}%
\newcommand{\xmark}{\ding{55}}%
\usepackage{fancyvrb}
\usepackage{url}
\usepackage{listings}
\newcounter{lstmain}
\usepackage{graphicx,xcolor,enumitem}
\usepackage{booktabs}

\lstnewenvironment{code}[1][]
{ \vspace{0.3cm}\footnotesize{\textsc{Code Listing \thelstmain: #1}}
  \hspace{0.1cm} \hrulefill
  \lstset{language=C++, basicstyle=\ttfamily\scriptsize,
    keywordstyle=\color{blue}\bfseries,commentstyle=\color{mygreen},
    stringstyle=\color{red}
  }
}
{
  \hrule \vspace{0.3cm}
  \addtocounter{lstmain}{1}
}

\lstnewenvironment{codeln}[1][]
{\textbf{Code Listing} \hspace{1cm} \hrulefill \lstset{language=C++, basicstyle=\ttfamily\scriptsize, numbers=left, numberstyle=\tiny, stepnumber=1, numbersep=5pt, keywordstyle=\color{blue}\bfseries,commentstyle=\color{mygreen}, stringstyle=\color{red}}}
{\hrule\smallskip}

\lstnewenvironment{smallcode}[1][]
{\lstset{language=C++, basicstyle=\ttfamily\scriptsize, keywordstyle=\color{myblue}\bfseries,commentstyle=\color{mygreen}, stringstyle=\color{red}}}
{\smallskip}

\xdefinecolor{mygreen}{RGB}{0,220,0}
\xdefinecolor{myblue}{RGB}{26,150,255}

\begin{document}

\title{A Parallel General Purpose Multi-Objective Optimization Framework,
  with Application To Electron Beam Dynamics}

\author{N. Neveu}
\altaffiliation[Also at ]{Argonne National Laboratory, USA}

\author{L. Spentzouris}
\affiliation{Illinois Institute of Technology, Chicago, IL}

\author{A. Adelmann}
\email{andreas.adelmann@psi.ch}
\author{Y. Ineichen }
\author{A. Kolano}
\altaffiliation[Also at ]{
	University of Huddersfield, West Yorkshire, United Kingdom and  CERN, Genf}
\author{C. Metzger-Kraus}
\affiliation{
	PSI, Villigen, Switzerland}%

\author{C. Bekas}
\author{A. Curioni}

\affiliation{IBM Research, Zurich, Switzerland }%

\author{P. Arbenz}
\affiliation{%
	Department of Computer Science, ETH Zurich, Switzerland}%

\date{\today}

\begin{abstract}
Particle accelerators are invaluable tools for research in the basic and applied sciences, such as materials science, chemistry,
  the biosciences, particle physics, nuclear physics and medicine. The design, commissioning, and operation of accelerator facilities is a
  non-trivial task, due to the large number of control parameters and the complex interplay of several conflicting design goals.
  The Argonne Wakefield Accelerator facility has some unique challenges resulting from its purpose to carry out advanced accelerator R\&D.
  Individual experiments often have challenging beam requirements, and the physical configuration of the beamlines is often changed
  to accommodate the variety of supported experiments. The need for rapid deployment of different operational settings
  further complicates the optimization work that must be done for multiple constraints and challenging operational regimes. 
  One example of this is an independently staged two-beam acceleration experiment which requires the construction 
  of an additional beamline (this is now in progress).  The high charge drive beam, well into the space charge regime, must be threaded
  through small aperture (17.6 mm) decelerating structures.  
  In addition, the bunch length must be sufficiently short to maximize power generation in the decelerator.  
We propose to tackle this problem by means of multi-objective optimization algorithms which also facilitate a parallel deployment.
In order to compute solutions in a meaningful time frame, a fast and scalable software framework is required.
In this paper, we present a general-purpose framework for simulation-based
multi-objective optimization methods that allows the automatic investigation of optimal sets of machine parameters.
Using evolutionary algorithms as the optimizer and \textsc{OPAL} as the forward solver, validation experiments and 
 results of multi-objective optimization problems in the domain of beam dynamics are presented. 
  Optimized solutions for the new high charge drive beamline found by the framework were used to finish the design 
  of a two beam acceleration experiment.
  The selected solution along with the associated beam parameters is presented.
  
\end{abstract}

\maketitle

\section{INTRODUCTION} \label{sec:introduction}

Particle accelerators play a significant role in many aspects of science and
  technology.
Fields, such as material science, chemistry, the biosciences, particle
  physics, nuclear physics and medicine depend on reliable and effective
  particle accelerators, both as research and practical tools.
Achieving the required performance in the design, commissioning, and operation
  of accelerator facilities is a complex and versatile problem.
Despite the success of on-line models in some facilities~\cite{xiaobiao}, 
and various model dependent and model independent tuning and optimization techniques, 
empirical tuning by operators is a common method used at many research facilities.
When the beam dynamics is nonlinear, as is the case with space charge, 
simple and fast models are applicable only in a very restricted manner. 
This further complicates any mult-objective optimization by complicating the model. 
In order to be able to reliably identify optimal configurations of
  accelerators, we solve large multi-objective design optimization
  problems to automate the investigation for an optimal set of tuning parameters.
  This approach has been used in the past with much success~\cite{hofler13,bazarov05,yrss:09,jefferson,gull1,gull2,marija}.
The difference here being the implementation and application to a 
problem at the Argonne Wakefield Accelerator Facility (AWA).

A hallmark of the AWA facility is the flexibility to swap physical
components in the beamlines, which enables different, often novel,
accelerator research experiments to take place.  Not only do the
physical machine components change, the beam characteristics also vary
considerably to meet different needs.  The facility operates at both low
and high charge (up to 100 nC), and at high charge strong nonlinearities
require a full 3D space charge approach in simulations.  
Finding optics solutions in this regime, especially when
there are additional constraints such as the small aperture two-beam
accelerating structures, is challenging even without the quick
turnaround of the beamline configurations.  Therefore, it has been an
important research objective to develop a precise, e.g. 3D model
embedded into a multi-objective optimization framework that may be used
as a flexible platform for optimization of changing machine
configurations operated at different charge levels. While other codes,
such as GPT~\cite{gpt} and ELEGANT~\cite{elegant}, also have
integrated genetic optimization algorithms; the OPAL~\cite{opal} framework
differentiates itself by being open source (i.e. free to use), massively
parallel, and fully 3D.

\begin{figure}
	\center
\includegraphics[width=0.5\textwidth]{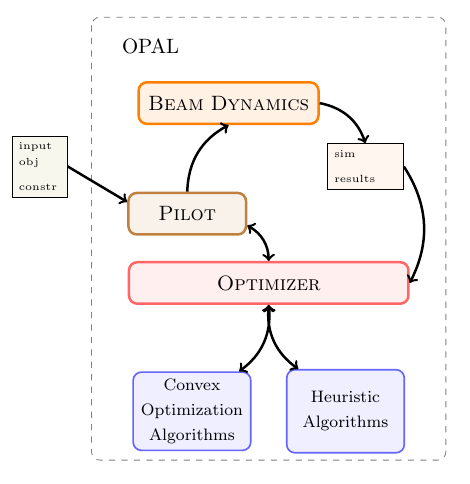}
\caption{Multi-objective framework: the pilot (master) solves the
	optimization problem specified in the input file by coordinating optimizer
	algorithms and workers running forward solves.}
\label{fig:framenetwork}
\end{figure}

A modular multi-objective software framework was developed (see
 Fig.~\ref{fig:framenetwork}) where the core functionality of the optimizer is decoupled from
 the ``beam dynamics'' but fully integrated in the OPAL framework. 
To that end, we use a master/slave mechanism where a master process governs a
 set of slave processes given some computational tasks (beam dynamics simulation) to complete.
This separation allows easy interchange of optimization algorithms, forward
  solvers and optimization problems.
A ``pilot'' coordinates all efforts between the optimization algorithm and the
  beam dynamics task. In the following sections, we will also use the notion of ``forward solver'' to indicate the beam dynamics task.
This forms a robust and general framework for massively parallel
  multi-objective optimization.
Currently the framework offers one concrete optimization algorithm, an
  evolutionary algorithm employing a \textsc{NSGA-II} selector \cite{pisa}.
Normally, simulation based approaches are plagued by the trade off between
  level of detail and time to solution.
This problem is addressed later in Section~(\ref{awa:subsection:test}) by using forward solvers with different time and
  resolution complexity.

The framework discussed here, incorporates the following three contributions:
\begin{enumerate}
  \item Implementation of a scalable optimization algorithm capable of
        approximating Pareto fronts in high dimensional spaces,
  \item design and implementation of a modular framework that is simple to use
        and deploy on large scale computational resources, and
  \item demonstration of the usefulness of the proposed framework on a real world
        application in the domain of particle accelerators. This is done
        with the optimization problem set as 
        the high charge photoinjector at the AWA. 
\end{enumerate}

The next section introduces the notation of multi-objective optimization
theory and describes the first implemented optimizer.
In Section~\ref{sec:framework}, the implementation of the framework is discussed.
We introduce the employed forward-solver in Section~\ref{sec:forward-solver}.
A validation and a proof of concept application in the beam dynamics problems 
mentioned above is discussed in Section~\ref{sec:experiments}.

\section{MULTI-OBJECTIVE OPTIMIZATION} \label{sec:optimization}

Optimization problems deal with finding one or more feasible solutions
  corresponding to extreme values of objectives.
If more than one objective is present in the optimization problem this is called
  a multi-objective optimization problem (MOOP).
A MOOP is defined as
\begin{align}
  \text{ min} \quad & \quad f_m({\bf x}), ~& m &= 1, \dots, M, \label{eq:moop:obj}\\
  \text{s.t.} \quad & \quad g_j({\bf x}) \geq 0, & j &= 0, \dots, J,
  \label{eq:moop:constr}\\
  \quad & \quad  x_i^L \leq {\bf x}=x_i \leq x_i^U,& i &=0, \dots, n
  \label{eq:moop:dvar} \text{,}
\end{align}
where $\bf f$ denotes the objectives (\ref{eq:moop:obj}),
  $\bf g$ the constraints (\ref{eq:moop:constr}),
  and $\bf x$ the design variables (\ref{eq:moop:dvar}).
Often, conflicting objectives are encountered, and this complicates the concept of
  optimality. \textit{Pareto optimality} is often used in such situations.
The set of Pareto optimal points forms the Pareto front or
  surface.
All points on this surface are considered to be Pareto optimal.
Sampling Pareto fronts is far from trivial.
A number of approaches have been proposed,
  e.g.\ evolutionary algorithms~\cite{deb:09},
  simulated annealing~\cite{kigv:83},
  swarm methods~\cite{keeb:95},
  and many more~\cite{domc:96,cati:02,kara:05,hoss:09}.
In the next section, we briefly introduce the theory of evolutionary algorithms
  used in the present work.

\subsection{Evolutionary Algorithms}

Evolutionary algorithms are loosely based on nature's evolutionary
  principles to guide a population of individuals towards an improved solution
  by honoring the ``survival of the fittest'' principle.
This ``simulated'' evolutionary process preserves entropy (or diversity in
  biological terms) by applying genetic operators, such as mutation and
  crossover, to remix the fittest individuals in a population.
Maintaining diversity is a crucial feature for the success of all evolutionary
  algorithms.

In general, a generic evolutionary algorithm consists of the following
  components:
\begin{itemize}
  \item \textit{Genes}: traits defining an individual,
  \item \textit{Fitness}: a mapping from genes to a fitness value for each
    individual,
  \item \textit{Selector}: selecting the $k$ fittest individuals of a
    population based on some sort of ordering,
  \item \textit{Variator}: recombination (mutations and crossover) operators
    for offspring generation.
\end{itemize}

\begin{figure}
    \centering
    \includegraphics[width=0.6\textwidth]{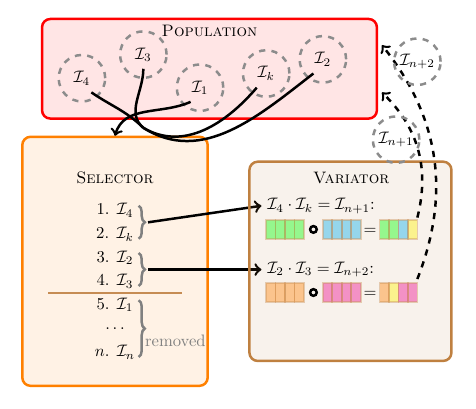}
  \caption{Schematic view of interplay between selector and variator. The
  selector ranks all individuals in the population according to fitness and
  subsequently the variator uses the fittest individuals to produces new
  offspring. Finally, the new children are reintroduced in the population.}
  \label{fig:varsel}
\end{figure}

Applied to multi-objective optimization problems, genes correspond to
  design variables.
The fitness of an individual is loosely related to the value of the objective
  functions for the corresponding genes.
Figure~\ref{fig:varsel} schematically depicts the connection of the
  components introduced above.
The process starts with an initially random population of individuals, each
  individual with a unique set of genes and corresponding fitness,
  representing one location in the search space.
In the next step, the population is processed by the selector
  determining the $k$ fittest individuals according to their fitness values.
While the $k$ fittest individuals are passed to the variator, the
  remaining $n-k$ individuals are eliminated from the population.
The \textsc{Variator} mates and recombines the $k$ fittest individuals to
  generate new offspring.
After evaluating the fitness of all the freshly born individuals a
  \textit{generation} cycle has completed and the process can start anew.

Since there already exist plenty of implementations of evolutionary algorithms,
  it was decided to incorporate the PISA library \cite{pisa} into our
  framework.
One of the advantages of PISA is that it separates variator from selector,
  rendering the library expandable and configurable.
Implementing a variator was enough to use PISA in our framework and
  retain access to all available PISA selectors.
As shown in Fig.~\ref{fig:varsel}, the selector is in charge of ordering a
  set of $d$-dimensional vectors and selecting the $k$ fittest individuals
  currently in the population.
The performance of a selector depends on the number of objectives and the
  surface of the search space.
So far, the NSGA-II selector \cite{dpam:02} has been used and exhibits satisfactory
  convergence performance.

The task of the variator is to generate offspring and ensure diversity in the
  population.
The variator can start generating offspring once the fitness of every
  individual of the population has been evaluated.
This explicit synchronization point defines an obvious bottleneck for parallel
  implementations of evolutionary algorithms.
In the worst case, some MPI processes are taking a long time to compute the
  fitness of the last individual in the pool of individuals to evaluate.
During this time all other resources are idle and wait for the result of
  this one individual in order to continue to generate and evaluate offspring.
To counteract this effect, the selector is already called when two individuals
  have finished evaluating their fitness, lifting the boundaries between
  generations and evaluating the performance of individuals.
New offspring will be generated and MPI processes can immediately return to
  work on the next fitness evaluation.
By calling the selector more frequently (already after two offspring
  individuals have been evaluated) results in better populations since bad
  solutions are rejected earlier.
On the other hand, calling the selector more often is computationally more
  expensive. Note this capability is also present in GPT's~\cite{gpt} optimization system. 

The variator implementation uses the master/slave architecture, presented in
  the next section, to run as many function evaluations as possible in parallel.
Additionally, various crossover and mutation policies are available for tuning
  the algorithm to the optimization problem.

\section{THE FRAMEWORK} \label{sec:framework}

Simulation based multi-objective optimization problems are omnipresent in
  research and industry.
These simulation and optimization problems are in
  general very big and computationally demanding.
This motivated us to design a massively parallel general purpose framework.
The key traits of such a design can be summarized as:
\begin{itemize}
  \item support any multi-objective optimization method,
  \item support any function evaluator: simulation code or measurements,
  \item offer a general description/specification of objectives, constraints
        and design variables,
  \item run efficiently in parallel on current large-scale high-end clusters
        and supercomputers.
\end{itemize}

\subsection{Related Work}

Several similar frameworks, e.g.~\cite{fide:09,lems:09,lbjt:07,dnld:06}, have
  been proposed.
Commonly these frameworks are tightly coupled to an optimization algorithm,
  e.g.\ only providing evolutionary algorithms as optimizers.
Users can specify optimization problems, but cannot change the
  optimization algorithm.
Our framework follows a more general approach, providing a user-friendly way
  to introduce new or choose from existing built-in multi-objective
  optimization algorithms.
Tailoring the optimization algorithm to the optimization problem at hand is
  an important feature due to the many different characteristics of
  optimization problems that should be handled by such a general framework.
As an example, it is shown how \textsc{Pisa}~\cite{pisa}, an existing evolutionary
  algorithm library, was integrated with ease.
Similarly, other multi-objective algorithms could be incorporated and
  used to solve optimization problems.

The framework presented in \cite{lems:09} resembles our implementation the
  most, aside from their tight coupling with an evolutionary algorithm
  optimization strategy.
The authors propose a plug-in based framework employing an island
  parallelization model, where multiple populations are evaluated concurrently
  and independently up to a point where some number of individuals of the
  population are exchanged.
This is especially useful to prevent the search algorithm to get stuck in
  a local minimum.
A set of default plug-ins for genetic operators, selectors and other
  components of the algorithms are provided by their framework.
User-based plug-ins can be incorporated into the framework by implementing a
  simple set of functions.

Additionally, as with simulation based multi-objective optimization, we can
  exploit the fact that both the optimizer and simulation part of the process
  use a certain amount of resources.
The ratio of work between optimizer and simulation costs can be reflected in
  the ratio of number of processors assigned to each task.
This not only provides users with  great flexibility in using any simulation
  or optimizer, but renders influencing the role assignment easy as well.

\subsection{Components}

\begin{figure}
  \centering
  \includegraphics[width=0.7\linewidth]{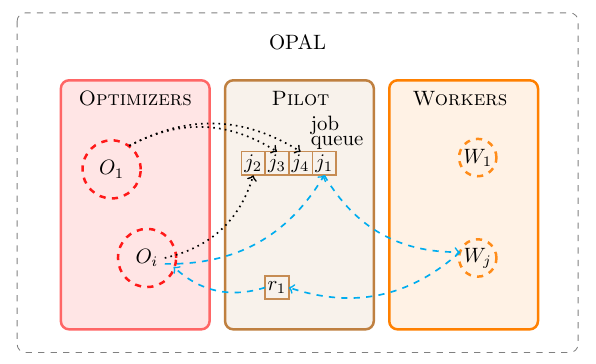}
  \caption{Schematic view of messages passed within the network between the
    three roles.
  The dashed cyan path describes a request (job $j_1$) sent from $O_i$ to the
  \textsc{Pilot} being handled by $W_j$. Subsequently the result $r_k$ is
  returned to the requesting \textsc{Optimizer} ($O_i$). The work $W_j$ are beam dynamics 
  simulation within OPAL.}
  \label{fig:opt-framework-layout}
\end{figure}

The basic assumption in simulation-based optimization is that a
  call to an expensive simulation software component present in the
  constraints or objectives is needed.
The framework is divided in three exchangeable components, as shown in
  Fig.~\ref{fig:opt-framework-layout}, to encapsulate the major behavioral
  patterns of the framework.

The \textsc{Pilot} component acts as a bridge between the optimizer and
  forward solvers, providing the necessary functionality to handle passing
  requests and results between the \textsc{Optimizer} and the
  \textsc{Simulation} modules.
The framework was implemented in \texttt{C++}, utilizing features like template
parameters to specify the composition of the framework.
``Default'' implementations are provided that can be controlled via command line options.
Due to its modular design, all components can be completely customized.

Every available MPI process will take up one of the three available roles (see
  Fig.~\ref{fig:framenetwork}):  one process acts as \textsc{Pilot}, the
  remaining processes are divided amongst \textsc{Worker} and
  \textsc{Optimizer} roles.
Both, the \textsc{Worker} and the \textsc{Optimizer} can consist of multiple
  MPI processes to exploit parallelism.
As shown in Fig.~\ref{fig:opt-framework-layout}, the \textsc{Pilot} is used
  to coordinate all ``information requests'' between the \textsc{Optimizer}
  and the \textsc{Worker}.
An information request is a job that consists of a set of design variables
  (e.g.~the genes of an individual) and a type of information it requests
  (e.g.~function evaluation or derivative).
The \textsc{Pilot} keeps checking for idle \textsc{Worker} and assigns jobs
  in the queue to any free \textsc{Worker}.
Once the \textsc{Worker} has computed and evaluated the request its results
  are returned to the \textsc{Optimizer} that originally requested the
  information.

After a process gets appointed a role, it starts a polling loop to asynchronously
  check for appropriate incoming requests.
To that end a \textsc{Poller} interface helper class has been introduced.
The \textsc{Poller} interface consists of an infinite loop that checks
  periodically for new MPI messages.
Upon reception a new message is immediately forwarded to the appropriate
  handler: the \texttt{onMessage()} method.
The method is called with the \texttt{MPI\_Status} of the received message and
  a \texttt{size\_t} value specifying different values depending on the value
  of the \texttt{MPI\_Tag}.
The \textsc{Poller} interface allows the implementation of special methods
  (denoted \textit{hooks}) determining the behavior of the polling process,
  e.g.\ for actions that need to be taken after a message has been handled.
Every \textsc{Poller} terminates the loop upon receiving a special MPI tag.

\subsection{Implementing an Optimizer}

All \textsc{Optimizer} implementations have to respect the API shown in
Listing 2.

\begin{code}[Optimizer API]
virtual void initialize() = 0;

// Poller hooks
virtual void setupPoll() = 0;
virtual void prePoll() = 0;
virtual void postPoll() = 0;
virtual void onStop() = 0;
virtual bool onMessage(MPI_Status status,
                       size_t length) = 0;
\end{code}

All processors running an \textsc{Optimizer} component call the
  \texttt{initialize} entry point after role assignment in the
  \textsc{Pilot}.
The implementation of \texttt{initialize} must set up and start the poller and
  the optimization code.
Since an optimizer derives from the \texttt{Poller} interface, predefined
  hooks can be used to determine the polling procedure.
Hooks can be implemented as empty methods, but the \texttt{onMessage}
  implementation should reflect the optimization part of the protocol for
  handling events from the \textsc{Pilot}.
A special set of communicator groups serves as communication channels to the
  \textsc{Pilot}, its job queue, and processes supporting the
  \textsc{Optimizer} component.

\subsection{Implementing a Forward Solver}

In most cases, forward solver implementations are simple wrappers to run
  an existing ``external'' simulation code using a set of design variables as
  input. In the case of the OPAL integration, the \texttt{main} function is
  playing the role of the ``forward solver''. To underline the general nature of our approach, 
  in a similar project, the described methods are used for cavity shape optimisation based on \cite{ARBENZ2008381}. 
As for the \textsc{Optimizer} component there exists a base class, labeled
  \texttt{Simulation} as common basis for all \textsc{Simulation}
  implementations.
In addition, this component also inherits from the \texttt{Worker} class,
  already implementing the polling protocol for default worker types.
As shown in the API in Listing 3, the \texttt{Worker} class expects an
  implementation to provide implementations for those three methods.

\begin{code}[Simulation API]
virtual void run() = 0;
virtual void collectResults() = 0;
virtual reqVarContainer_t getResults() = 0;
\end{code}

First, upon receiving a new job, the \texttt{Worker} will call the \texttt{run} 
method on the \textsc{Simulation} implementation.
This expects the \textsc{Simulation} implementation to run the simulation in a 
\textit{blocking} fashion, meaning the method call blocks and does not return
until the simulation has terminated.
Subsequently, the \texttt{Worker} calls \texttt{collectResults}, where the
\textsc{Simulation} prepares the result data, e.g. parsing output files,
and stores the requested information in a \texttt{reqVarContainer\_t} data structure.
Finally, the results obtained with \texttt{getResults} are sent to the \textsc{Pilot}. 
As before, the serialized data is exchanged using MPI point-to-point communication using a specific set of communicators.

\subsection{Specifying the Optimization Problem}

We aimed at an easy and expressive way for users to specify multi-objective optimization problems.
Following the principle of keeping metadata (optimization and simulation input data) together, 
we decided to embed the optimization problem specification in the simulation input file by 
prefixing it with special characters, e.g. as annotations prefixed with a special character.
In some cases, it might not be possible to annotate the simulation input file.
By providing an extra input file parser, optimization problems can be read from stand-alone files.
To allow arbitrary constraints and objective expressions, such as
%
\begin{Verbatim}[fontsize=\scriptsize]
  name: OBJECTIVE,
        EXPR="5 * average(42.0, "measurement.dat") + ENERGY";
\end{Verbatim}
%
\noindent
An expression parser using Boost Spirit~\cite{boost} was implemented.
In addition to the parser, we need an evaluator able to evaluate an expression,
given a parse tree and variable assignments to an actual value.
Expressions arising in multi-objective optimization problems usually evaluate
to booleans or floating point values.
The parse tree, also denoted abstract syntax tree (AST), is constructed recursively while an expression is parsed.
Upon evaluation, all unknown variables are replaced with values, 
either obtained from simulation results or provided by other subtrees in the AST.
In this stage, the AST can be evaluated bottom-up and the desired result is
  returned after processing the root of the tree.

To improve the expressive power of objectives and constraints, a
  simple mechanism to define and call custom functions in expressions was introduced.
Using simple functors, to compute an
  average over a set of data points, enriches expressions with custom
  functions.
Custom function implementations overload the \texttt{()} parenthesis operator.
The function arguments specified in the corresponding expression are stored in
  a \texttt{std::vector} of Boost variants~\cite{boost2} that can be
  booleans, strings or floating point values.

All custom functions are registered with expression objects.
This is necessary to ensure that expressions know how they can resolve
  function calls in their AST.
As shown in Listing 5 this is done by creating a collection of Boost
  functions~\cite{boost3} corresponding to the
  available custom functions in expressions and passing this to the
  \textsc{Pilot}.

\begin{code}[Creating function pointer for registering functor]
functionDictionary_t funcs;
client::function::type ff;
ff = average();
funcs.insert(std::pair<std::string, 
		client::function::type> 
       		("my_average_name", ff));
\end{code}

A set of default operators, corresponding to a mapping to \texttt{C} math
  functions, is included in the dictionary by default.
This enables an out of source description of optimization problems containing
  only simple math primitives.

\subsection{Parallelization} \label{sec:parallelization}

The parallelization is defined by a mapping of the roles introduced above to
  available cores.
Command-line options allow the user to steer the number of processors used in
  worker and optimizer groups.
Here, we mainly use the command-line options to steer the number of processors
  running a forward solver.

One major disadvantage of the master/slave implementation model is the fast
  saturation of the network links surrounding the master node.
In \cite{bctg:09} authors observe an exponential increase in hot-spot latency
  with increasing number of workers that are attached to one master process.
The limiting factor is the number of outgoing links of a node in the
  network topology. For a few workers, the links surrounding a
  master process are subject to congestion.
This effect is amplified further by large message sizes.

To that end we implemented a solution propagation based on rumor networks 
(see \cite{bgps:06,ayss:09}) using only one-sided communication.
This limits the number of messages sent over the already heavily used links
surrounding the master node and helps to prevent the use of
global communication. Using information about the interconnection network topology and the
application communication graph, the task of assigning roles helps to further
improve the parallel performance.

\section{FORWARD SOLVER} \label{sec:forward-solver}

The framework contains a wrapper implementing the API mentioned in
  Listing 3 for using \textsc{OPAL}~\cite{opal} as the forward solver.
\textsc{OPAL} provides different trackers for cyclotrons and linear
  accelerators with satisfactory parallel performance. 
With access to the \textsc{OPAL} forward solver, the framework is able to
  tackle a multitude of optimization problems arising in the domain of
  particle accelerators.
  The framework is also integrated into \textsc{OPAL} so that users can 
  define optimization problems within an input file, requiring no 
  additional knowledge or installation of the API to use it.


If the objectives and constraints are simple arithmetical expressions, 
the \texttt{FunctionEvaluator} simulator can be used.
Using functors and the default expression primitives, 
multi-objective optimization problems can be specified, 
i.e.\ the benchmark problem presented in \cite{hbwh:05}:

	\begin{widetext}
		\begin{align} \label{eqn:bench}
		\text{min} & \left[ 1 - \exp \left( -1 \left(
		\left(x_1 - \frac{1}{\sqrt{3}} \right)^2 +
		\left(x_2 - \frac{1}{\sqrt{3}} \right)^2 +
		\left(x_3 - \frac{1}{\sqrt{3}} \right)^2 \right)\right), \right. \\
		\vspace{3em} 
		& \left. 1 - \exp \left( -1 \left(
		\left(x_1 + \frac{1}{\sqrt{3}} \right)^2 +
		\left(x_2 + \frac{1}{\sqrt{3}} \right)^2 +
		\left(x_3 + \frac{1}{\sqrt{3}} \right)^2 \right)\right) \right]^T \nonumber \\
		\vspace{3em} \nonumber \\
		\text{s.t.} \quad & \quad -1 \le x_i \le 1, \quad i=1,2,3 \nonumber
		\text{.} \nonumber
		\end{align}
	\end{widetext}

\section{EXPERIMENTS} \label{sec:experiments}

In this section numerical results of the validation benchmark and optimization 
of a photoinjector operated in the space charge dominated regime is presented.

\subsection{Optimizer Validation}

\begin{figure}
    \centering
      \includegraphics[width=0.4\linewidth]{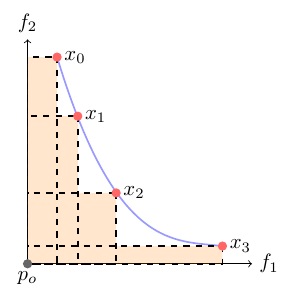}
  \caption{The hypervolume for a two-objective optimization problem
  corresponds to the shaded area formed by the dashed rectangles spanned by
  all points on the Pareto front and an arbitrary selected origin $p_o$.}
  \label{fig:hypervolume}
\end{figure}

\begin{figure}
  \centering
    \includegraphics[width=0.6\linewidth]{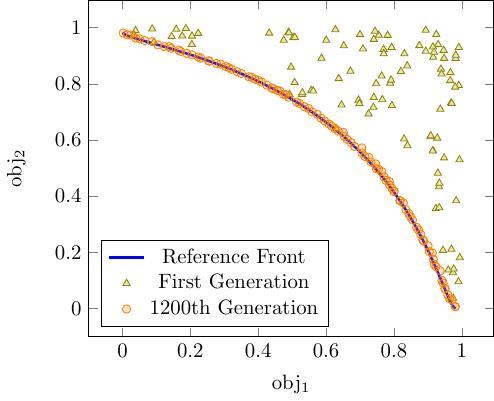}
  \caption{Variator benchmark after $1100$ function evaluations using binary
           crossover and independent gene mutations (each gene mutates with
           probability $p=\frac{1}{2}$) on a population of $100$
           individuals.}
  \label{fig:pisa_bench}
\end{figure}

To ensure that the optimizer works correctly, the benchmark
  problem (\ref{eqn:bench}) was solved.
To that end, we use a metric for comparing the quality of a Pareto front.
Given a point in the Pareto set, we compute the $m$ dimensional volume (for
  $m$ objectives) of the dominated space, relative to a chosen origin.
This is visualized for $2$ objectives in Fig.~\ref{fig:hypervolume}.
For further information and details of the implementation see~\cite{whbb:12}.
Figure~\ref{fig:pisa_bench} and the corresponding hypervolume values in
  Table~\ref{tbl:bench_rms_error} show expected convergence.
The reference Pareto front is clearly very well approximated.
It took a total of 1100 function evaluations to perform this computation.
The hypervolume of the reference solution ($0.6575$) for our benchmark was
 computed by sampling the solution provided in~\cite{hbwh:05}.
\begin{table}
\begin{center}
  \caption{Convergence of benchmark problem with errors relative to
    hypervolume of sampled reference solution.}
  \label{tbl:bench_rms_error}
  \begin{tabular}{c|c|c}
	\cline{1-3}
    tot.\ function  & hypervolume & relative error\\
    evaluations    & & \\
	\cline{1-3}
    100  &  0.859753 & $3.076 \times 10^{-1}$ \\
    200  &  0.784943 & $1.938 \times 10^{-1}$ \\
    500  &  0.685183 & $4.210 \times 10^{-2}$ \\
    900  &  0.661898 & $6.689 \times 10^{-3}$ \\
    1100 &  0.657615 & $1.749 \times 10^{-4}$ \\
	\cline{1-3}
  \end{tabular}
\end{center}
\end{table}
Table~\ref{tbl:bench_rms_error} shows satisfactory
  convergence to the sampled reference Pareto front after 1000 (plus the
  additional 100 evaluations for the initial population) function evaluations.

\subsection{AWA Photoinjector Optimization} \label{awaproblem}
Next the optimization framework is applied to the high charge beam line
 at the Argonne Wakefield Accelerator (AWA) facility. 
The goal of this optimization is to produce beams of electrons that meet 
design specifications; this includes number of particles (charge), energy, 
and particle distribution (characterized by beam sizes and energy spread).
As shown in Fig.~\ref{awa-linac}, the installed portion of the the 
beam line consists of an rf photocathode gun, 
two solenoids, and six linear accelerating cavities
followed by four quadrupoles and a stripline kicker. 
The charge of interest, 40 nC, is needed for two beam acceleration (TBA) 
experiments performed at the AWA~\cite{gai_power_jing_2012,JING201872}, 
which motivates this work. 
Prior experimental results were limited by beam size when the beam passed through small aperture 
wakefield structures located downstream. 
In an attempt to maximize charge transmission in upcoming experiments,
magnet strengths of the solenoids and quadrupoles leading into the 
TBA section of the beam line were optimized, shown in Fig.~\ref{awa-tba}.
The simulation model includes from the gun to the septum.
The optimization location is chosen as the  
entrance to the first quadrupole on the dog leg ($s_3$), see Fig.~\ref{awa-tba}.
Minimizing beam sizes here will enable capture and further focusing before space charge effects dominate the beam. 
This will also enable cleaner transport through downstream elements.

\begin{figure*}
\includegraphics[width=0.9\linewidth]{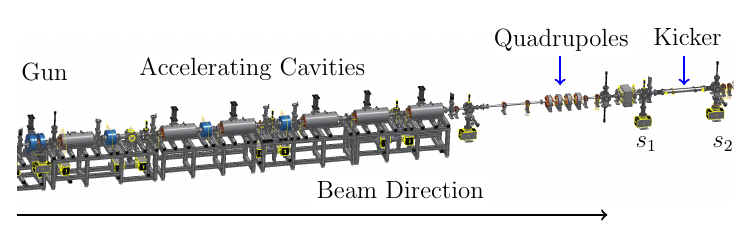}
	\caption{Side view of the high charge linac at the AWA. 
		All hardware in this drawing is currently installed. 
	Note locations $s_1$ and $s_2$, before and after the kicker.}
	\label{awa-linac}
\end{figure*}

\begin{figure*}
\includegraphics[width=0.9\linewidth]{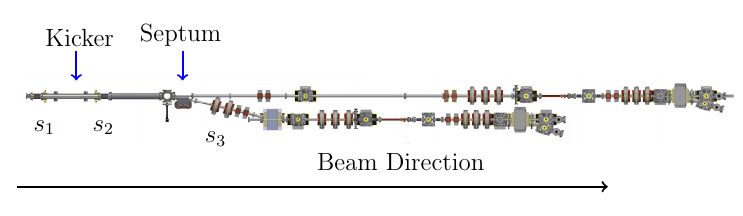}
	\caption{Continuation of the high charge beam line layout at the AWA, top view. 
		This is the proposed two beam acceleration section. 
		Only the kicker in this drawing is installed. 
	Note $s_3$, the entrance to the fifth quadrupole on the beam line.
	This is the optimization location.}
	\label{awa-tba}
\end{figure*}
%

In addition to addressing the challenge of producing an optimized beam, 
this model was chosen to demonstrate the ability of the framework to tackle large problems.
Six design variables and objectives were used, along with three constraints.
The objectives include transverse and longitudinal beam sizes, 
transverse momentum, and longitudinal energy spread. 
The design variables include the two gun solenoids and 
the first four quadrupoles strengths. 
This problem encompasses high dimensionality 
and nonlinear effects such as space charge. 
Using this model it was possible to find good solutions that meet the 
operational goals for the new beamline.  One of these solutions is 
presented in the next section.

\subsubsection{Time Step Scan} \label{awa:subsection:test}
\begin{figure}
	\centering
	\includegraphics[width=0.8\linewidth]{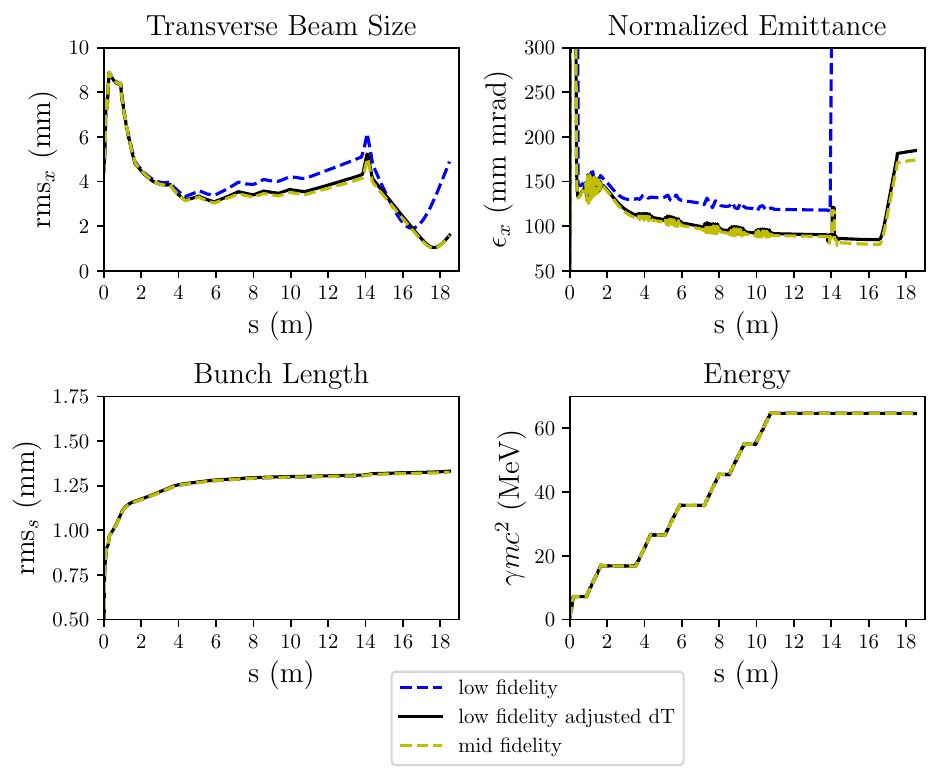}
	\caption{Comparison of different fidelity models (dT stands for time step).}
	\label{tstep}
\end{figure}

Before running a full scale optimization of the problem described in Subsection \ref{awaproblem}, 
a study on time step and number of particles in the simulation model 
was done to reduce the time of the simulation while 
maintaining the physics of interest. 
The grid size $16 \times 16 \times 32$ was chosen, 
and parallelized in the x and y directions.
After comparing several options (1,000, 10,000, 20,000, 50,000, 100,000) 
with a small time step, the number of particles was fixed at 10,000.
Next several time steps were explored, see Table~\ref{timestep}.
The largest steps were too big to resolve the beam parameters accurately.
See low fidelity plot in Fig.~\ref{tstep} for $dT=$\num{5E-11} results.  
\begin{table}
	\begin{center}
		\caption{Checkmarks (\cmark) indicate desired beam parameters are resolved at that time step. 
			An (\xmark) indicates the time step is too large, and results are nonphysical.}
		\label{timestep}
		\begin{tabular*}{0.48\textwidth}{c|c|c|c}
	\cline{1-4}
			Time Step, dT (s) \quad & Linac & Drift & Quadrupoles \\
	\cline{1-4}
			$5 \times10^{-10}$  & \xmark & \xmark & \xmark \\
			$1 \times10^{-10}$  & \xmark & \xmark & \xmark \\
			$5 \times10^{-11}$  & \xmark & \xmark & \xmark \\
			$1 \times10^{-11}$  & \cmark & \cmark & \xmark \\
			$5 \times10^{-12}$  & \cmark & \cmark & \xmark \\
			$1 \times10^{-12}$  & \cmark & \cmark & \cmark \\
	\cline{1-4}
		\end{tabular*}
	\end{center}
\end{table}

In the drifts and linac tanks, $dT=\num{1E-11}$ was sufficient. 
However, it was not acceptable near the quadrupoles. 
For all models, the longitudinal parameters (rms$_s$ and energy) 
are calculated correctly, but discrepancies are seen in the transverse 
(rms$_x$ and $\epsilon_x$) for low fidelity results. This discrepancy is 
what led to the decision to adjust the time steps w.r.t. beam line elements. 
In the linac and drift sections $dT=\num{1E-11}$ was used. 
Near sensitive elements such as the quadrupoles, kicker, and septum, 
a time step of $dT=\num{1E-12}$ was used.
The resulting simulations are low fidelity in most places, but closely approximate 
the mid fidelity simulations for metrics of interest, as shown in 
Fig.~\ref{tstep}.  Mid fidelity simulations used steps of  $dT=\num{1E-12}$ everywhere.
The average run time of each simulation with the adjusted time steps was 1.6~minutes.
In comparison, the mid fidelity simulation ran for 18~minutes.
Note, a smaller time step, \num{1E-13}, is always used in the gun where the 
beam has low energy and is changing rapidly.

\subsubsection{Hyper parameter Scan}
While the optimization problem and goals were well defined (Subsection \ref{awaproblem}), 
it was not clear what the best hyper parameters for the genetic algorithm would be.
These parameters include gene mutation probability, mutation probability, 
recombination probability, number of individuals, 
and number of generations to complete. 
Given the beam line in Fig.~\ref{awa-linac},
four small optimization experiments were done with various hyper parameters. 
Similar to the time step scan, 
the goal of this exercise was to determine which set of optimization
parameters strongly influence the results, 
and whether there was a time to solution difference.
From here on, we will reference each experiment as ex-1, ex-2, ex-3, and ex-4
as shown in Table \ref{extable}. 
\begin{table}
\begin{center}
\caption{Input Parameters for initial twenty four hour AWA optimization experiments. 
The gene mutation probability was equal to the mutation probability (not shown) in all four experiments. 
The max number of individuals per generation was~80.}
\label{extable}
\begin{tabular}{c|c|c|c}
\cline{1-4}
& Gene Mutation \qquad& Recombination \qquad & Number of completed \vspace{-0.75em}\\
Experiment & Probability & Probability &  generations \\ 
\cline{1-4}
ex-1 &  0.1  & 0.9  &  96 \\
ex-2 &  0.3  & 0.7  &  81 \\
ex-3 &  0.8  & 0.2  &  53 \\
ex-4 &  0.01 & 0.09 &  95 \\ 
\cline{1-4}
\end{tabular}
\end{center}
\end{table}

The maximum number of individuals per generation was fixed at 80. 
This number was chosen based on the node architecture, and the 
to prevent a prohibitive computational cost.  
Each experiment was allowed to run for twenty four hours, with 
a maximum generation limit of 100. 
We reduced the six objectives to four, 
and shortened the simulation time by moving the objectives further 
upstream to $s_1$ and $s_2$, the locations before and after the kicker, 
see Fig.~\ref{awa-tba}.  
The objectives include: $\varepsilon_{x}\left(s = s_1\right)\text{, } \varepsilon_{x}\left(s = s_2\right)$, $\text{rms}_{s}\left(s = s_1\right)\text{, and }  \text{rms}_{s}\left(s = s_2\right)$. 
The OPAL input file for these cases and all subsequent optimization runs in this paper, 
can be found at in following repository: \url{https://github.com/nneveu/awa-tba}.

\begin{figure}
	\centering
	\begin{minipage}{0.49\textwidth}
		\centering
		\includegraphics[width=\textwidth]{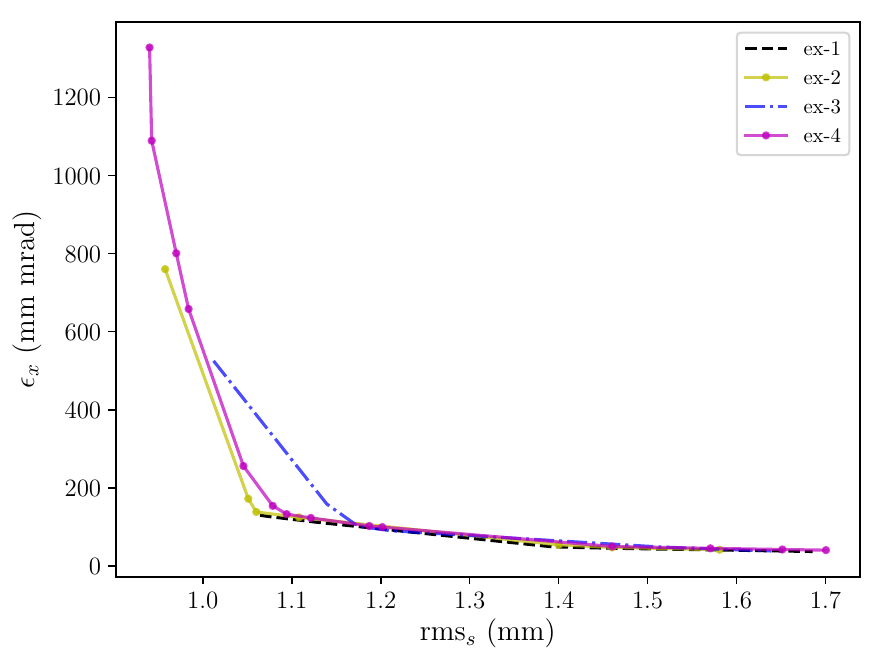}
		(a) Pareto fronts for ex-1 through ex-4 at $s_1$.
	\end{minipage}
	\begin{minipage}{0.49\textwidth}
		\centering
		\includegraphics[width=\textwidth]{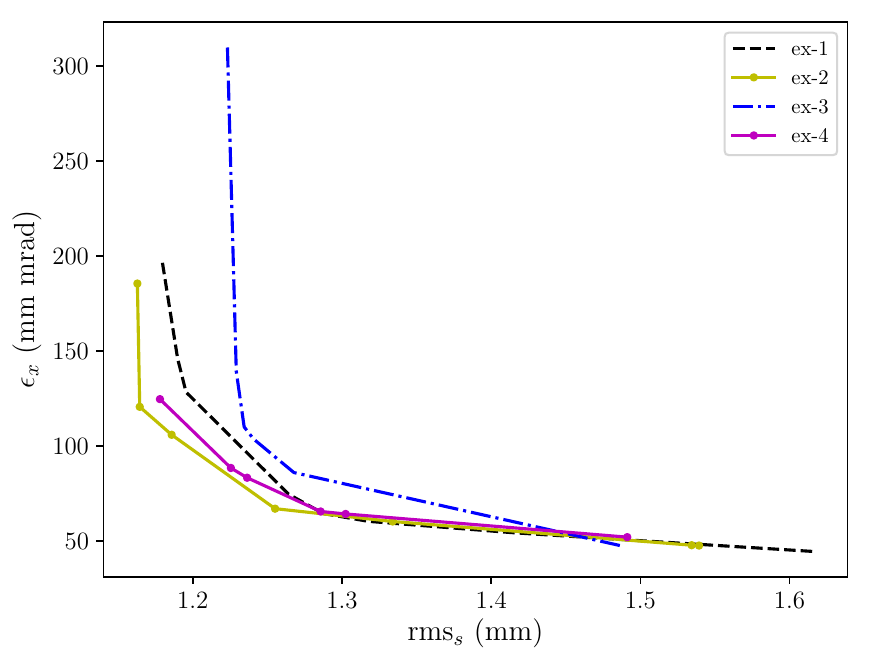}
		(b) Pareto fronts for ex-1 through ex-4 at $s_2$.
	\end{minipage}
	\caption{Comparison of Pareto fronts for initial optimization experiments, ex-1 through ex-4.}
	\label{expareto}
\end{figure}

After collection of the data for all four experiments, several metrics
were compared, including number of generations completed in twenty four hours and
Pareto fronts at $s_1$ and $s_2$, see Fig.~\ref{awa-linac}.
From Table \ref{extable}, we clearly see ex-3 is significantly 
slower, as it evaluated only 53 generations 
compared to the experiment with the maximum number, ex-1 at 96 generations.
Perhaps this trade off would be acceptable if the Pareto front was significantly 
improved, but from Fig. \ref{expareto}, but this is not the case.
Similar arguments can be made for ex-2, which evaluated about 15 less generations.
The Pareto fronts at $s_2$, are nearly identical. It is expected
this trend would continue given more time. 
When looking at the Pareto front at $s_1$, only ex-4 has a slightly 
larger range compared to the others.
With ex-2 and ex-3 eliminated due to evaluation time, 
and a slightly better Pareto front at $s_1$ for ex-4, 
the hyper parameters in ex-4 were chosen as the default values for subsequent runs.

\subsubsection{TBA Optimization Problem}
With computational and hyper parameters set, 
the optimization problem of interest is explored.
The objectives (beam sizes and energy spread) are calculated at 
$s_3=19.4$~m, located downstream of the septum, see Fig.~\ref{awa-tba}. 
Given the longitudinal location of $s_3$ (unless otherwise noted), 
we define the objectives and input parameters as:
\begin{align}
\text{min}  \quad & \text{rms}_{x}, \quad \text{rms}_{y} \label{eq:awa:p1}\\
& \text{rms}_{px}, \quad \text{rms}_{py}, \label{eq:awa:p2}\\
& \text{rms}_{s}, \quad \text{dE} \label{eq:awa:p4} \\
\text{constraints} \quad & \text{rms}_x < 0.1 \, (\text{m}) |_{s=s_1}\label{eq:awa:c1}\\
\quad & \text{rms}_y < 0.1\, (m) |_{s=s_1}\, \label{eq:awa:c2}\\
\quad & |\text{rms}_y - \text{rms}_x | < 0.005 \, (\text{m}) |_{s=s_1}\label{eq:awa:c3}\\
\text{subject to} \quad & q = 40 \, (\text{nC}) \label{eq:awa:firstconstr}\\
\quad & \text{Volt}_{\text{Gun}} = 64 \, \, (\text{MV/m}) \label{eq:awa:lastconstr}\\
\quad & \text{Volt}_{\text{Linac}} = 24 \, \text{or} \, 25\,\, (\text{MV/m}) \\
\quad & R_x = R_y = 9 \, (\text{mm}) \label{eq:awa:firstdvar}\\
\quad & \phi_{\text{gun}} =-20^\circ \label{eq:awa:gphidvar}\\
\quad & \phi_{\text{linac}} =-20^\circ \label{eq:awa:lastdvar}
\end{align}

The first four objectives, parameters (\ref{eq:awa:p1}) to (\ref{eq:awa:p2}),
minimize the transverse ($rms_{x,y}$) beam size and transverse momentum ($rms_{px,py}$)
at the location of interest in the beam line ($s_3$, see Fig.~\ref{awa-tba}). 
Minimizing the beam size at this location is essential to 
to preventing loss of particles by scraping; 
which ensures better transmission through the wakefield structures downstream.
These structures are called Power Extractor and Transfer Structure (PETS), 
and will be located after the septum and quadrupoles in Fig.~\ref{awa-tba}.
This device is putting the tightest constraints on the beam parameters.  
The aperture diameter is \SI{17.6}{mm} and the bunch length should be below \SI{2}{mm}
to facilitate large power extraction~\cite{PETSeq}.   
Less divergence in the beam (lower transverse momentum spread) 
reduces growth of transverse beam size after the focal point (location of min beam size).
This reduces halo by ensuring the beam is not over focused through a hard waist.
The momentum spread is also critical to preventing large growth during transport. 
All of these factors help with transmission downstream. 

The next two objectives in parameter (\ref{eq:awa:p4}) minimize the 
longitudinal beam size ($rms_s$), and energy spread (dE) at location $s_3$. 
This helps reduce the transverse beam size growth in bending elements.
A small bunch length ($rms_s$) is also critical to the goals of 
TBA experiments. The power generated in the wakefield structures 
designed for TBA is related to the bunch length \cite{JING201872,PETSeq}.
Eqs.~\ref{eq:awa:c1} to \ref{eq:awa:c3} 
define three constraints used to guide the algorithm.
However, it is important to not over-constrain the problem, which would prevent
the algorithm from converging.
The difference constraint, Eq.~\ref{eq:awa:c3}, is used to favor nearly round beams.
This prevents one dimension from becoming disproportionately large compared to the other.
At the AWA, there is some room in the beam pipe to allow the y dimension to grow, but round beams are preferred.
Equations (\ref{eq:awa:firstconstr}) to
(\ref{eq:awa:lastdvar}) define the charge, gun voltage, linac voltages, 
laser radius, gun phase, and linac cavity phases (in that order). 
These are parameters in the simulation that must be defined, but do not vary during the optimization.
For setup of the AWA design variables, objectives, and constraints in the OPAL input file, 
refer to the the repository above.

Design variables include the currents in two gun solenoids (IBF and IM), 
and four quadrupole strengths (KQ1-KQ4). The objectives include
beam size (transverse and longitudinal), transverse momentum, and energy spread as
defined in Eqs. (\ref{eq:awa:p1}) to (\ref{eq:awa:p4}). 
The location at the entrance of the kicker is $s_1=16.45$~(m), 
and the objectives are optimized at location $s_3=19.4$~(m). 
This is the entrance to the fifth quad in the beam line. 
This location is where the beam should be captured and focused through subsequent elements.

\subsubsection{AWA Optimization Results}
All simulations for this experiment were carried out on Bebop a
high performance computing (HPC)
cluster provided by the Laboratory Computing Resource Center (LCRC)
at Argonne National Laboratory (ANL). Intel Knights Landing 
(KNL) processors at 1.3 GHz with 128 GB of memory 
and 64 cores per node were used for all runs. 
There are 352 compute nodes available on 
Bebop, with a total of 22,528 cores. All jobs were run and compared 
on 8 cores each, which allowed 8 jobs per node on the KNLs.
This in combination with the number of nodes available 
allows for very large optimization jobs, like the AWA case.
Typical runs for this paper used 41 nodes, which corresponds to 2624 KNL cores 
and a generation size of 328 individuals.

With the time steps and hyper parameters set by the work in Section \ref{awa:subsection:test}, 
the optimization problem described in \ref{awaproblem} was run for 200 generations.
The initial number of individuals was fixed at 656, 
and the minimum number individuals in later generations was fixed at 328. 
These numbers were in part based on the architecture of the KNL's. 
Since each simulation takes 8 cores, and there are 64 cores per KNL node, 
a large population size that would fit evenly on these resources was chosen. 
Again, the location of optimization is $s_3=19.4$~(m). 
\begin{figure}
	\begin{center}		
		\includegraphics[width=0.65\textwidth]{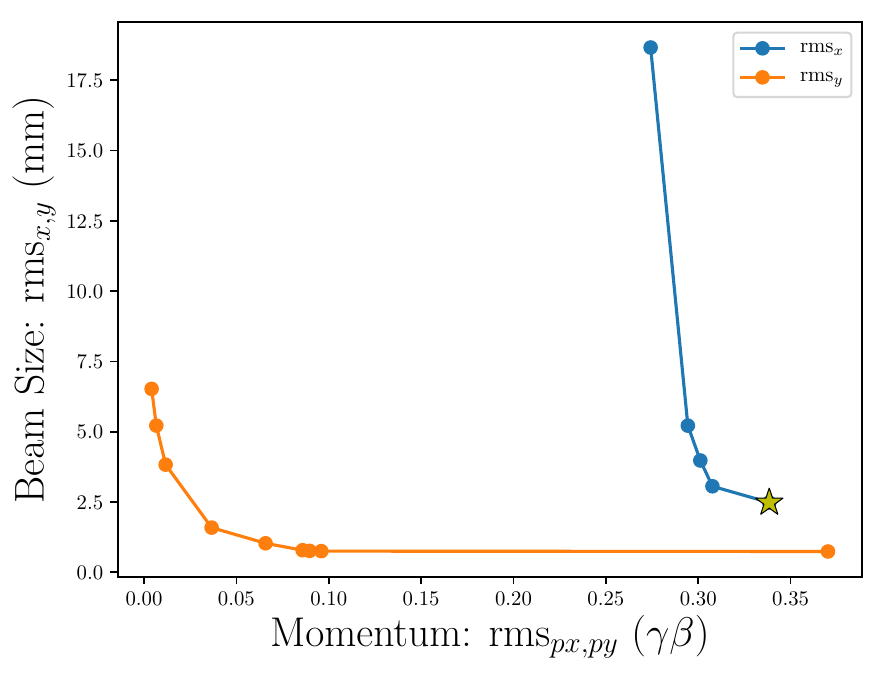}
	\end{center} 
		\caption{Pareto front comparing transverse beam sizes (rms$_{x,y}$) and transverse momentum (rms$_{px,py}$). The yellow star indicates the 
		point plotted in Fig.~\ref{fig:stat}}
	\label{fig:pareto1}
\end{figure}
 
\begin{figure}
	\includegraphics[width=0.65\textwidth]{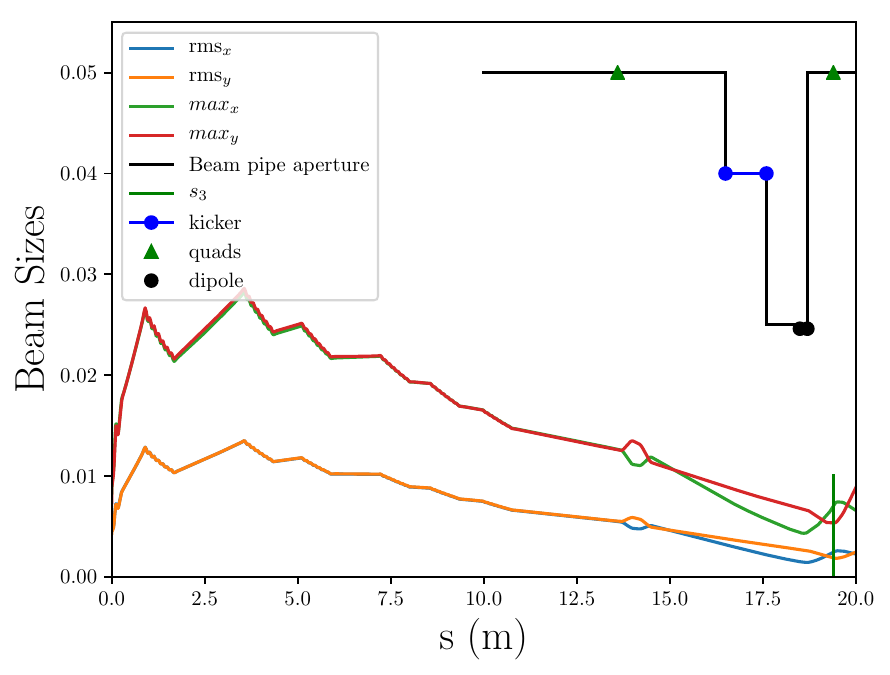}
	\caption{Optimized beam sizes along high charge beam line. The gun is located at $s=0$, 
	both x and y beam sizes are shown. The black line represents the relevant beam line aperture, while
	the green line indicates the location of the optimization.}
	\label{fig:stat}
\end{figure}

As expected, the x dimension is impacted by the bending elements, and unable to reach 
the small beam sizes seen in the y dimension. This suggests objectives in the x 
dimension will drive design variable choices used during operations. 
However, it is still necessary to 
include the y dimension in the optimization. Early optimization tests showed the y dimension 
can easily grow out of control if it is not included in the objectives.
Those results are not shown here due to the unfeasible nature of the solutions 
(i.e. $rms_y$ larger than the beam pipe).
In the case of bunch length, there are not many options to choose from, as the phase was not varied.
With these observations in mind, several beam parameters corresponding to
options on the Pareto Front in Fig.~\ref{fig:pareto1} were plotted and compared. 
A select result is shown in Fig.~\ref{fig:stat}.  
The maximum beam sizes are well below the beam pipe aperture limits, also shown in Fig.~\ref{fig:stat}.
The solution is nearly round, which will increase chances of keeping the beam nearly round
as it travels to the last triplet in Fig.~\ref{awa-tba}.
Overall this solution is satisfactory, and meets all requirements for the new TBA beamline at the AWA.

The TBA line under construction includes a kicker and septum combination to route one bunch train to its own PETS, 
while the other bunch train is undeflected and goes to a different PETS.  
The optimization results shown here start at the gun and go through to the first quadrupole after the septum ($s_3=19.4$~(m)).  
These results were used as the basis for the full solution up to the PETS~\cite{neveuthesis}.
After adjustments to include 3D rf field maps and CSR in the dipoles~\cite{neveuthesis}, 
the maximum beam sizes at the PETS location were below 10mm with a bunch length of rmsz = 1.6mm.  
Although the simulation results at the PETS location cannot yet be verified, 
since the beamline is not yet completely installed, some checks have been possible for the partially completed beamline. 
For example, beam images were taken on the YAG screen located downstream of the kicker, 
and it was found that the extracted beam sizes at different kicker angles 
were in good agreement with simulation results~\cite{neveuthesis}. 
When it becomes possible to test the complete simulation solution, 
it is expected that there will not be perfect agreement with simulation; 
for example there may be errors in the fields of beamline elements. 
However, the simulation can be a tool to track down the source of discrepancies, 
such as understanding the effect of the gun field on the beam symmetry. 
More importantly, the optimization results should provide significant improvement 
in efficiency for achieving a good operational regime.

\begin{table}
	\begin{center}
		\caption{Input Parameters for the optimized solution shown in Figures~\ref{fig:pareto1} and \ref{fig:stat}.}
		\label{tab:designopt}   
		\begin{tabular}{l|c|c}
			\cline{1-3}
			\textbf{Design Variable} \qquad \qquad & \textbf{Unit}	\quad&  \textbf{Value}  \\ 
			\cline{1-3}
			{Buck Focusing Solenoid} \qquad \qquad & amps	\quad & 478 \\
			Matching Solenoid &	amps	& 197	  \\
			Quadrupole 1& T-m		& -0.8	\\ 
			Quadrupole 2& T-m		& 0.9	\\
			Quadrupole 3 & T-m		& 0.8	\\
			Quadrupole 4 & T-m		& -1.0	\\ 
			Bunch Length & mm 		& 1.5	\\
			\cline{1-3}
		\end{tabular}
	\end{center}
\end{table}

\section{CONCLUSIONS} \label{sec:conclusions}

A general-purpose framework for solving multi-objective
  optimization problems was presented.
Its modular design simplifies the application to simulation-based optimization
  problems for a wide range of problems and allows to exchange the
  optimization algorithm.
The flexibility of being able to adapt both ends of the optimization
  process, the forward solver and the optimization algorithm simultaneously
  not only leads to broad applicability but it facilitates
  tailoring the optimization strategy to the optimization problem as well.

The framework was integrated into OPAL, and used 
to study a beam dynamics problem at the AWA.
A scan of time step and hyper parameters was done to determine computational settings.
Then a full scale physics optimization was performed.
Optimization of the 3D beam size and energy spread was accomplished.
The TBA beam line presented is currently being installed at the AWA.
Once installation is complete, the results shown here will guide future experiments at the AWA.

In contrast to approaches that are tightly coupled to the optimization
algorithm, the range of possible applications is much wider.
Even in cases where the mathematical model of the forward solver is not known
exactly, fixed or real time measurements can be used to guide the
search for the Pareto optimal solutions.
Finally, combining a multi-objective optimization framework, such as 
the one presented, with practical experience in the field should expedite 
the decision making process in the design and operation of particle accelerators.

\section{ACKNOWLEDGMENT}

The authors thank the AWA team for contributing to the
  formulation of optimization problems. 
  We gratefully acknowledge the computing resources provided on Bebop,
  a HPC cluster operated by the LCRC at ANL.
  Thanks to Scott Doran for providing CAD drawings of the AWA beam lines.
  This work was partly supported by the 
  U.S. Department of Energy, Office of Science, under 
  contract number DE-AC02-06CH11357 and grant number DE-SC0015479.

\providecommand{\SortNoop}[1]{}
\end{document}